\documentstyle[multicol,aps,psfig,epsfig,bbox]{revtex} 
\begin{document} 
\newcommand{\vk}{{\vec k}} 
\newcommand{\vK}{{\vec K}}  
\newcommand{\vb}{{\vec b}}  
\newcommand{{\vp}}{{\vecp}}  
\newcommand{{\vq}}{{\vec q}}  
\newcommand{\vQ}{{\vec Q}} 
\newcommand{\vx}{{\vec x}} 
\newcommand{\vh}{{\hat{v}}} 
\newcommand{\tr}{{{\rm Tr}}}  
\newcommand{\beq}{\begin{equation}} 
\newcommand{\eeq}[1]{\label{#1} \end{equation}}  
\newcommand{\half}{{\textstyle\frac{1}{2}}}  
\newcommand{\gton}{\stackrel{>}{\sim}} 
\newcommand{\lton}{\mathrel{\lower.9ex 
\hbox{$\stackrel{\displaystyle <}{\sim}$}}}  
\newcommand{\ee}{\end{equation}} 
\newcommand{\ben}{\begin{enumerate}}  
\newcommand{\een}{\end{enumerate}} 
\newcommand{\bit}{\begin{itemize}}  
\newcommand{\eit}{\end{itemize}} 
\newcommand{\bc}{\begin{center}}  
\newcommand{\ec}{\end{center}} 
\newcommand{\bea}{\begin{eqnarray}}  
\newcommand{\eea}{\end{eqnarray}} 
\newcommand{\beqar}{\begin{eqnarray}}  
\newcommand{\eeqar}[1]{\label{#1} 
\end{eqnarray}}

\title{High-$p_T$ Tomography of $d+Au$ and $Au+Au$  at SPS, RHIC, and LHC}

\author{Ivan~Vitev$^{*,\dagger}$ and Miklos~Gyulassy$^{*}$ \\[2ex] }

\address{ $^*$Department of Physics, Columbia University,  
        538 West 120-th Street, New York, NY 10027 \\[.5ex] 
        $^\dagger$Department of Physics and Astronomy, 
        Iowa State University, Ames, IA 50011} 
 
\maketitle

\begin{abstract} 
The interplay of nuclear effects on the $p_T > 2$~GeV inclusive 
hadron spectra  in $d+Au$ and $Au+Au$ reactions at 
$\sqrt{s}_{NN} = 17, 200, 5500$~GeV is compared  to 
leading order perturbative QCD calculations  
for elementary $p+p\, (\bar{p}+p)$ collisions. 
The competition between nuclear shadowing,  
Cronin effect, and jet energy loss due to  
medium-induced gluon radiation is predicted to lead to a striking 
energy dependence of the nuclear suppression/enhancement pattern 
in $A+A$ reactions. We show that future $d+Au$ data can used to
disentangle the initial and final state effects.
\vspace{.2cm} 
 
\end{abstract}

\begin{multicols}{2} 
 
{\em Introduction.} Tomography is the study of the properties of  
matter through the attenuation pattern of fast particles that  
propagate and lose energy as a result of multiple elastic and  
inelastic scatterings. Recently, this technique has been applied in the 
field of nuclear physics~\cite{VECTOMO,TOMO} to map out the evolution of  
the QCD matter density produced in ultra-relativistic heavy ion reaction. 
It is based on the theoretical advances in understanding QCD multi-parton  
interactions  in non-Abelian media~\cite{ELOSS,GLV}.

The determination of the opacity of the transient    
quark-gluon plasma (QGP) produced in such reactions 
via jet tomography requires  
theoretical control over the interplay between many competing  
nuclear effects that modify the high-$p_T$ hadron spectra. 
These include nuclear modifications to the parton distribution
functions (PDFs), referred to as shadowing~\cite{SHADOVER},   
Cronin effect~\cite{CRON},  and jet quenching~\cite{HIJING}, as well 
as the energy dependence of the underlying pQCD  parton spectra. 
In this   letter we propose an approach to disentangle these effects 
by comparing  high-$p_T$ hadron yields 
in  $p+p\, (\bar{p}+p)$, $d+A\, (p+A)$, and $A+A$ reactions over a very  
wide energy range.  
In particular, predictions  are presented for $d+Au$ and central  
$Au+Au$  at center of mass energies per nucleon  
$\sqrt{s}_{NN}=17, 200, 5500$~GeV typical of the CERN 
Super Proton Synchrotron (SPS), the Relativistic Heavy Ion Collider (RHIC), 
and the future Large Hadron Collider (LHC).

We demonstrate that at SPS the puzzling absence of quenching of  
$\pi^0$ in central $Pb+Pb$~\cite{NOELOSS} can be understood as 
due to a  larger than previously expected Cronin enhancement~\cite{GLVel}  
dominating over our  predicted~\cite{GLV} suppression due to jet energy  
loss. At RHIC energies, on the other hand,  we find that   
quenching dominates over both Cronin and shadowing effects. 
Furthermore, the interplay of these effects leads to  
a surprising approximately constant suppression factor
of the Glauber geometry scaled~\cite{GLAU} pQCD prediction
in the  $4 \leq p_T \leq 20$~GeV range.  
At LHC we predict that the $\pi^0$ suppression factor  
is substantially larger than at RHIC but also decreases systematically
with transverse momentum in the $6 \leq  p_T \leq 100$~GeV  range due to 
the higher initial gluon densities expected and the hardening of 
the underlying initial jet spectra.

{\em  Particle spectra in $d+A$ and $A+A$.}  The scaling of   
high-$p_T$ hadron production in  $d+A$ and $A+A$ is simply 
controlled by nuclear geometry in the absence of initial and 
final state interactions. The Glauber multiple collision  
model~\cite{GLAU} can be used to calculate  the number of 
binary nucleon-nucleon collisions at any  impact 
parameter $b$. In $p+A$ the experimental {\em cross section}  
has usually been presented  without centrality selection 
while in $A+A$  reactions { \em hadron  
multiplicity distributions} are generally presented 
with restricted centrality (impact parameter $b$) cuts. 
Dynamical nuclear effects for these cases are detectable through 
the nuclear modification ratio  
\begin{equation}         
R_{BA}(p_T)  = \left\{   
\begin{array}{ll}  \displaystyle 
\frac{d \sigma^{dA}}{dyd^2{\bf p}_T} / 
\frac{ 2A \; d \sigma^{pp}} {dyd^2{\bf p}_T} \;\; &  
{\rm in}  \; \;    d+A  \\[2.8ex]  \displaystyle 
\frac{d N^{AA}(b)}{dyd^2{\bf p}_T} / 
\frac{ T_{AA}(b)\; d \sigma^{pp}}{dyd^2{\bf p}_T}
   \;\;  &    {\rm in} \;  \; A+A  \; 
\end{array} \right. ,  
\label{geomfact} 
\end{equation}          
where $2A$ and $T_{AA}({b}) = 
\int  d^2{\bf r} \;T_A({\bf r})T_B({\bf r}-{\bf b})$ 
in terms  of nuclear thickness functions 
$T_A(r)=\int dz \;\rho_A({\bf r},z)$ 
are the corresponding Glauber scaling factors of $d\sigma^{pp}$.  
The  lowest order pQCD differential cross section for  inclusive 
$A+B\rightarrow h+X$ production that enters Eq.(\ref{geomfact})
is given by 
\begin{eqnarray} 
\left.   
\begin{array}{r}  \!   \displaystyle   \frac{1}{2A}  
\frac{d\sigma^{dA}}{dyd^2{\bf p}_T} \\[2.8ex]  
 \displaystyle  \frac{1}{T_{AA}(b)} 
\frac{d N^{AA}(b)}{dyd^2{\bf p}_T} 
\end{array} \!\! \right\} \!  &=&  
K \!  \sum_{abcd} \int\! \!dx_a  
dx_b  \! \int \!  d^2 {\bf{k}}_a d^2{\bf{k}}_b \,   
g({\bf{k}}_a) g({\bf{k}}_b)    \nonumber \\[-1ex] 
&\;&  \times \; S_A(x_a,Q^2_a)  S_B(x_b,Q^2_b)  \nonumber \\[.5ex] 
&\;& \times \; f_{a/A}(x_a,Q^2_a) f_{b/B}(x_b,Q^2_b)  
\,  \frac{d\sigma}{d{\hat t}}^{ab\rightarrow cd}  \nonumber \\[.5ex] 
&\;& \times \int_0^1 d \epsilon \, P(\epsilon) 
 \frac{z^*_c}{ z_c}  \frac{D_{h/c}(z^*_c,{Q}_c^2)}{\pi z_c} \; . 
\label{hcrossec} 
\end{eqnarray} 
In Eq.(\ref{hcrossec}) $x_a, x_b$ are the initial momentum fractions  
carried by the hard-scattered partons with probabilities sampled from the  
PDFs $f_{\alpha/A}(x_\alpha,Q_\alpha^2)$. 
The momentum fraction carried away by the leading hadron $z_c=p_h/p_c$  
is sampled from the fragmentation functions (FFs) $D_{h/c}(z_c,Q_c^2)$.  
We use the leading order (LO) Gl\"{u}ck-Reya-Vogt (GRV98) parameterization  
of PDFs~\cite{SFUNCS} and the LO Binnewies-Kniehl-Kramer (BKK) 
parameterization of the FFs~\cite{FFUNCS}.

A constant  $K$-factor and parton $k_T$ broadening function,   
$g({\bf k})= \exp(-{\bf k}_T^2 /\langle {\bf k}_T^2 \rangle_{pp}) / 
 \pi\langle {\bf k}_T^2 \rangle_{pp} $,  are included  
to account phenomenologically for next-to-leading  
order corrections. 
In systematic fits~\cite{KARIPP} to the inclusive  
hadron production in $p+p\, (\bar{p}+p)$  reactions 
these parameters are fixed from data~\cite{CRON,PPDATA}. 
The $K$-factor drops out in the ratios of $B+A$ to scaled $p+p$, 
but the phenomenological $k_T$ broadening in $p+p$  
is essential to establish an accurate nuclear geometry scaled baseline. 
We find that a fixed $\langle k_T^2\rangle_{pp}=1.8 \;{\rm GeV}^2$ 
reproduces to within 30\% the  spectral shapes in $p+p \, (p+\bar{p})$ 
for $p_T>2$~GeV over the whole energy range of interest. 
 
In $B+A$ reactions isospin effects are accounted  
for on average in the PDFs for a nucleus with $Z$ protons and 
$N$ neutrons  via  $ f_{\alpha/A}(x_\alpha,Q_\alpha^2)=  
(Z/A) \, f_{\alpha/p}(x_\alpha,Q_\alpha^2) 
 +  (N/A) \, f_{\alpha/n} (x_\alpha,Q_\alpha^2)$. 
Nuclear modifications to the PDFs~\cite{SHADOVER} are  
included through the shadowing function $S_A(x_\alpha,Q^2_\alpha)$ 
from the EKS98 parameterization~\cite{EKS98}.

The Cronin effect observed  in $p+A $  reactions relative to the 
Glauber-scaled $p+p$ result~\cite{CRON} is modeled  via  multiple 
initial state scatterings of the partons in cold nuclei.   
For an initial state parton  distribution $dN^{(0)}({\bf k}) $, 
random elastic  scattering induces further $k_T$-broadening
as shown for example in~\cite{GLVel}.  
The possibility of hard fluctuations along the projectile 
path leads to  a power law tail of the $k_T$ distribution  
that enhances  $\langle \Delta  k^2_T \rangle_\chi$ 
beyond the naive Gaussian random walk result ${\chi \mu^2 }$, where  
$\chi=\langle n \rangle = L/\lambda$ is the cold nuclear opacity  
in terms of the path length $L$ and 
the parton mean free path $\lambda$. The  screening scale  $\mu$   
regulates the Rutherford divergence in a cold nucleus. 
For $\chi \mu^2 \ll k_T^2 \le Q_{max}^2$
the Rutherford tail leads to a logarithmic enhancement  
of the mean square momentum transfer  
$\langle \Delta  k^2_T \rangle_\chi 
= \chi\mu^2 \ln( 1+ c \, Q_{max}^2/\mu^2)$, 
where $c$ depends on the detailed form of the kinematic cut-off. 
For a high energy parton with transverse  momentum $p_T$ produced
in a $p+A$ reaction $Q_{max}^2\sim p_T^2$. We therefore  model 
the Cronin effect by using 
\begin{equation} 
\langle k_T^2 \rangle_{pA}\approx  
\langle k_T^2 \rangle_{pp} + L_A \frac{\mu^2}{\lambda}   
\ln(1 +  c \, p_T^2/\mu^2 )  \;\; .  
\label{cron} 
\end{equation}  
in the $k_T$ broadening functions $g(k_a)$  
in Eq.(\ref{hcrossec}), taking $L_A=4/3 \, R_A$ 
as the mean nuclear thickness traversed. 
Fig.~1 shows that the calculation is consistent with the energy 
dependence $\sqrt{s} = 27.4, 38.8$~GeV~\cite{CRON} observed in 
$p+W/p+Be$  with transport parameters set as follows: 
$c/\mu^2=0.18 \, /{\rm GeV}^{2}$,  $\lambda=3.5$ fm and   
$\mu^2 / \lambda=0.05$~GeV$^2$/fm.  
These are consistent with   
$\mu^2/\lambda = 0.064\pm 0.036$~GeV$^2$/fm
extracted from  fits to Drell-Yan data~\cite{ARLEO}. 
Fig.~1 also shows that the expected Cronin + shadowing effect in
$p+W/p+Be$ at RHIC energies is much smaller than at lower energies
because the high-$p_T$ pQCD spectra  at $\sqrt{s}_{NN}=200$~GeV are 
considerably  less steep.

\begin{figure}[b] 
\hspace*{0.1cm }\epsfig{file=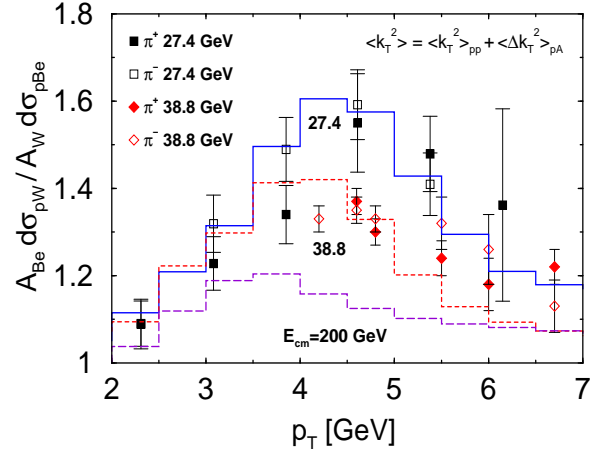,height=3.0in, 
width=2.2in,clip=5,angle=-90} 
\begin{minipage}[t]{8.6cm}  
\vspace*{.7cm} 
\caption{The ratio of $A$-scaled $p+W/p+Be$ data on $\pi^+,\pi^-$  
production at  $\sqrt{s}=27.4, 38.8$~GeV from~{\protect \cite{CRON}}. 
Calculations for $\half (\pi^+ + \pi^-$)  include  nuclear shadowing 
and initial  parton broadening as in Eq.(\ref{cron}) with  
$\mu^2 / \lambda = 0.05$~GeV$^2$/fm. The anticipated $\sqrt{s}=200$~GeV 
$p+W/p+Be$ ratio is also shown.} 
\end{minipage} 
\label{cron_wbe} 
\end{figure} 

The effects of multiple scattering and nuclear shadowing  
in $d+Au$ and $Au+Au$ {\em without} final state interactions 
at SPS, RHIC, and LHC  are shown in  
Fig.~2 for neutral pions. The numerical results 
for charged particles are comparable. Variations arise 
from the  different partonic contribution  and the 
correspondingly different shadowing for various 
hadron  species~\cite{KARIPP}. 
In our model of the Cronin effect, the enhancement in $Au+Au$  
at SPS energies of $\sqrt{s}_{NN}=17$~GeV  may reach  a factor 
$\sim 4$  at  $p_T \simeq 4-5$~GeV.  This is 
greater than observed in $Pb+Pb$ reactions and also 
greater than estimated with the Cronin model of Ref.~\cite{NOELOSS}. 
Unfortunately, at these low energies the results are extremely sensitive 
to model assumptions due the very rapid fall-off of the partonic spectra.  
We note that at least within our model, there is room for 
hadron suppression  due to energy loss even at SPS. 
At RHIC  for  $\sqrt{s}_{NN}=200$~GeV the Cronin enhancement   
spans the $p_T=1-8$~GeV range and is seen to peak at   
$p_T\simeq 3$~GeV.  Its maximum value in $d+Au$ and $Au+Au$ is 1.3 (1.6)   
respectively. Similar and even smaller magnitudes of the Cronin 
effect at RHIC have been  discussed in~\cite{OTHERCR}.  At higher 
transverse momenta the effects of isospin and shadowing lead to $R_{BA} 
\simeq 0.8$ at $p_T\simeq 20$~GeV.  At LHC energies of  
$\sqrt{s}_{NN}=5500$~GeV Cronin effect  is  overwhelmed   by  
shadowing at small $x$ ($p_T <$ 10~GeV) and anti-shadowing at larger $x$ 
when the EKS98~\cite{EKS98} parameterization is used.  
The net nuclear modification due to Cronin effect and shadowing at 
LHC is expected to be tiny ($ \leq 15\%$) throughout the $p_T$ range  
shown.

We turn next to the predicted  suppression effects in  
nucleus-nucleus reactions due to jet quenching~\cite{HIJING}.  
In Eq.(\ref{hcrossec}) this is taken into account in the 
fragmentation function via the modification of the momentum 
fraction carried away by the leading hadron. If a jet of momentum 
$p_c$ prior to hadronization looses a fraction $ 0\leq \epsilon < 1$ 
of its energy then  $z=p_h/p_c \rightarrow  z^*=z/(1-\epsilon)$.    
The distribution $P(\epsilon,E)$ of the fractional energy loss of 
a fast parton with energy $E$ due to multiple  gluon  emission is 
computed as in  Gyulassy-Levai-Vitev (GLV)~\cite{TOMO}.

\begin{figure}[thb] 
\hspace*{-0.4cm }\epsfig{file=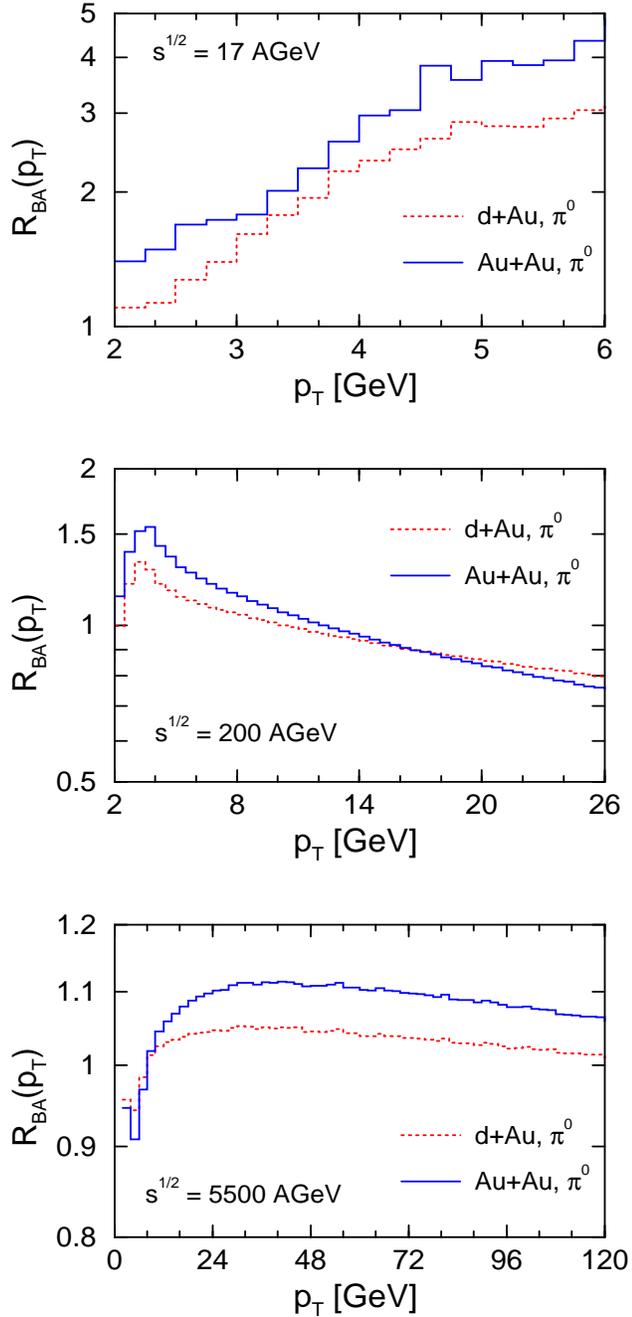,height=3.3in, 
width=7.0in,clip=5,angle=-90} 
\label{crondA} 
\begin{minipage}[t]{8.6cm}  
\vspace*{.2cm} 
\caption{The nuclear modification $R_{BA}(p_T)$   due to Cronin effect 
and shadowing (but not energy loss) for $\pi^0$  in $d+Au$  ($B=d,A=Au$) 
and central  $Au+Au$ ($B=A=Au$) reactions 
at $\sqrt{s}_{NN} =  17, 200, 5500$~GeV. } 
\end{minipage} 
\end{figure} 
We compute $P(\epsilon,E)$ taking into account the 
longitudinal Bjorken expansion  the plasma (gluon) density 
$\rho(\tau)= (\tau_0/\tau)   \rho(\tau_0)$, where 
$\tau_0 \rho_0= (1/\pi R_A^2) dN^g/dy$ relates to the gluon 
rapidity density produced in central $A+A$ that fixes the initial opacity. 
It has been shown that the azimuthally averaged  energy loss is insensitive  
to transverse expansion~\cite{V2}. 
The mean number of radiated gluons $\langle N^g(E) \rangle$ remains 
small due to the plasmon mass cut-off $\omega_{pl} \sim 0.5$~GeV~\cite{TOMO}.
Therefore, there is a finite $n=0$ (no radiation)  contribution 
$P_0(\epsilon,E) = e^{-\left\langle {N^g(E)}\right\rangle }\delta(\epsilon)$. 
We have checked the sensitivity of the results to  reducing the plasmon 
mass by a factor of two. This was found to lead to $\sim 25 \%$  more 
suppression  at $p_T=5$~GeV and to $ < 10\%$ increased suppression at 
$p_T=20$~GeV  for RHIC energies.

Our main results for central $Au+Au$  including all three nuclear 
effects  (Cronin+Shadowing+Quenching) are presented  in Fig.~3. 
Jet  tomography consists of determining the  effective initial gluon 
rapidity  density $dN^g/dy$ that best reproduces the quenching pattern
of the data~\cite{NOELOSS,PHEN,PHEN200,STAR200}. 
At SPS the large Cronin enhancement   
is reduced by a factor of two with $dN^g/dy = 350$ but the data
are more consistent with a smaller gluon density $ \lton 200$.
Unfortunately, as emphasized above, at this low energy the results are 
very sensitive to  the details of the model. At RHIC, for $p_T > 2$~GeV jet 
quenching  dominates, but surprisingly  the rate of variation with $p_T$ of 
the Cronin enhancement and jet quenching conspire to 
yield an approximately constant suppression pattern with magnitude dependent 
only on the initial $dN^g/dy$. At higher $p_T > 20$ GeV the softening
of the initial jet spectra due to the EMC modification of the   
PDFs compensates for the reduced energy loss. 
This unexpected interplay between the three nuclear effects at RHIC is 
the main prediction of this letter. 
At LHC energies the much larger gluon densities $dN^g/dy 
\sim 2000-3500$ are expected to lead to a  dramatic variation of 
quenching with $p_T$ as shown.

In nuclear media of high opacity the mean fractional energy loss 
$\langle \Delta E \rangle / E$ of moderately hard partons can become on 
the order of unity. For LHC this may be reflected in the 
$p_T \leq 10$~GeV region through deviations from the extrapolated 
high-$p_T$  suppression trend. Hadronic fragments coming from energetic jet   
would tend to compensate the rapidly increasing quenching (seen in Fig.~3)  
with decreasing transverse momentum and may restore the hydrodynamic-like 
participant scaling in the soft regime.

{\em  Conclusions.}  In this letter we predicted a characteristic 
evolution pattern of the {\em magnitude} and the $p_T$ {\em dependence} of
the nuclear modification factor in $d+A$ and $A+A$ reactions 
as a function of the center of mass energy per nucleon.  
A systematic approach was used to take into 
account Cronin effect, nuclear shadowing, as well as jet quenching. 
Our results suggest that at SPS energies  the  Cronin enhancement may be 
larger than expected previously,  leaving room for moderate energy loss. 
At RHIC we predict that the three nuclear effects in central $Au+Au$ 
lead to a surprising approximately constant suppression pattern of 
$\pi^0$ with $R_{AA}(p_T) \simeq 0.3-0.2$ for $dN^g/dy\sim 800-1200$.  
We emphasize that none of the nuclear effects alone would lead to
such a flat $R_{AA}(p_T)$.  At LHC shadowing and Cronin effect       
in the $6 \leq p_T \leq 100$~GeV range were found to be essentially negligible,
leading to  $\leq 15\%$ correction, while the jet quenching was predicted to 
be large and with a strong $p_T$ dependence.  
We emphasize the importance of future $d+Au$ data at RHIC to 
isolate and test the initial state 
Cronin and shadowing effects predicted  in Fig.~2. 
While it is still too early to draw conclusions 
from the preliminary data~\cite{PHEN200,STAR200} shown in Fig.~3, 
the combined future analysis
of $d+Au$ and $Au+Au$ high-$p_T$ measurements will improve the tomographic
determination of the initial gluon densities produced at RHIC.   
   
\begin{figure}[thb] 
\hspace*{-0.5cm }\epsfig{file=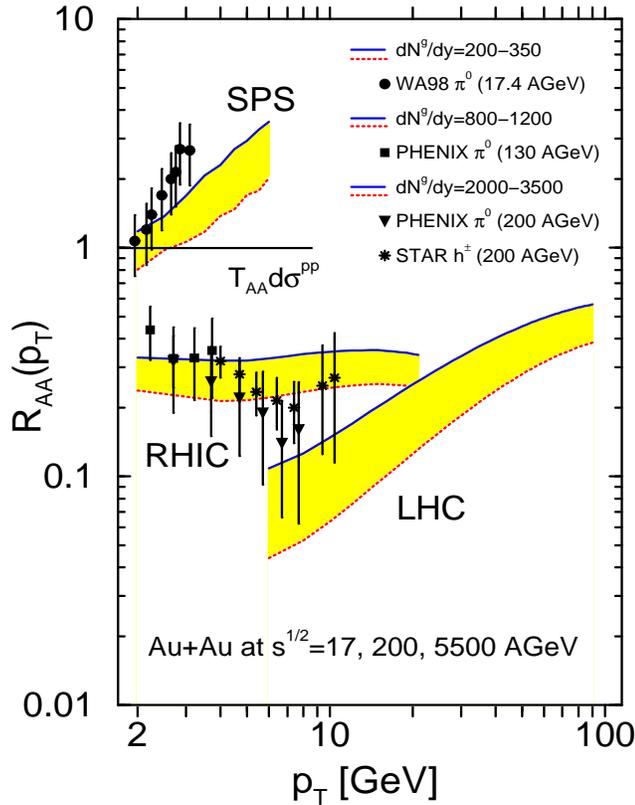,
height=3.3in,width=4.3in,clip=5,angle=-90} 
\begin{minipage}[t]{8.6cm}  
\vspace*{.2cm} 
\caption{The suppression/enhancement ratio $R_{AA}(p_T)$  
         ($A=B=Au$)   for neutral 
         pions at $\sqrt{s}_{NN}=17$, $200$, $5500$~GeV.   Solid (dashed) 
         lines correspond to the smaller (larger)  effective initial  
         gluon rapidity densities at given $\sqrt{s}$ that drive parton 
         energy loss. Data on
         $\pi^0$ production in central $Pb+Pb$ at $\sqrt{s}_{NN}=17.4$~GeV 
         from WA98~{\protect \cite{NOELOSS}} and in central $Au+Au$ at
         $\sqrt{s}_{NN}=130$~GeV~{\protect \cite{PHEN}}, as well as  
         {\em preliminary} data at 
         $200$~GeV ~{\protect \cite{PHEN200}}  from PHENIX and 
         $h^\pm$ central/peripheral data from 
         STAR~{\protect \cite{STAR200} are shown. The sum of 
         estimated statistical and systematic errors are indicated. } }  
\end{minipage} 
\label{sup_sys} 
\end{figure} 

{\em Acknowledgments:}  This work is  supported by the Director, 
Office of Science, Office of High Energy and Nuclear Physics, 
Division of Nuclear Physics, of the U.S. Department of Energy 
under Grant No. DE-FG02-93ER40764.

\end{multicols} 
\end{document}